\documentclass[preprint]{iucr}              
\usepackage{amsmath}
\usepackage{graphicx}
\usepackage{color}
\usepackage{hyperref}
     \journalcode{J}              

\begin{document}                  



\title{One-shot pair distribution functions of thin films using lab-based x-ray sources}


\author[a]{Johan}{Bylin}
\author[a]{Vassilios}{Kapaklis}
\cauthor[a]{Gunnar K}{P\'alsson}{gunnar.palsson@physics.uu.se}{\aff}

\aff[a]{Division of Materials Physics, Department of Physics and Astronomy, Uppsala University, Box 516, SE-75121, Uppsala, \country{Sweden}}

\maketitle     

\begin{synopsis}
Supply a synopsis of the paper for inclusion in the Table of Contents.
\end{synopsis}

\begin{abstract}
We demonstrate the feasibility of obtaining accurate pair distribution functions of thin amorphous films down to 80 nm, using modern laboratory-based x-ray sources. The pair distribution functions are obtained using a single diffraction scan (one-shot) without the requirement of additional scans of the substrate or of the air. By using a crystalline substrate combined with an oblique scattering geometry, most of the Bragg scattering of the substrate is avoided, rendering the substrate Compton scattering the primary contribution. By utilizing a discriminating energy filter, available in the latest generation of modern detectors, we demonstrate that the Compton intensity can further be reduced to negligible levels at higher wavevector values. We minimize scattering from the sample holder and the air by the systematic selection of pixels in the detector image based on the projected detection footprint of the sample and the use of a 3D printed sample holder. Finally, x-ray optical effects in the absorption factors and the ratios between the Compton intensity of the substrate and film are taken into account by using a theoretical tool that simulates the electric field inside the film and the substrate, which aids in planning both the sample design and measurement protocol. 
\end{abstract}


\section{Introduction}

Many thin film systems can exhibit exotic phases, whose thermal stability relies on the adhesion to the substrate, and/or the finite-size itself, and are therefore hard or impossible to synthesize in the bulk. Thin film techniques, owing to the fast quenching rate, allow access to a wider amorphous composition range than what is achievable with standard bulk techniques. Therefore, it is of great interest to be able to study the structural aspects of these thin films, in which obtaining their pair distribution function (PDF) is key. Thin film PDFs (tPDF) are also important in the sense that these can serve as an initial characterization and a starting point for more advanced studies at synchrotron facilities, and may lower the load on those instruments.

\begin{figure}
	\centering
	\includegraphics[width= 0.575\columnwidth]{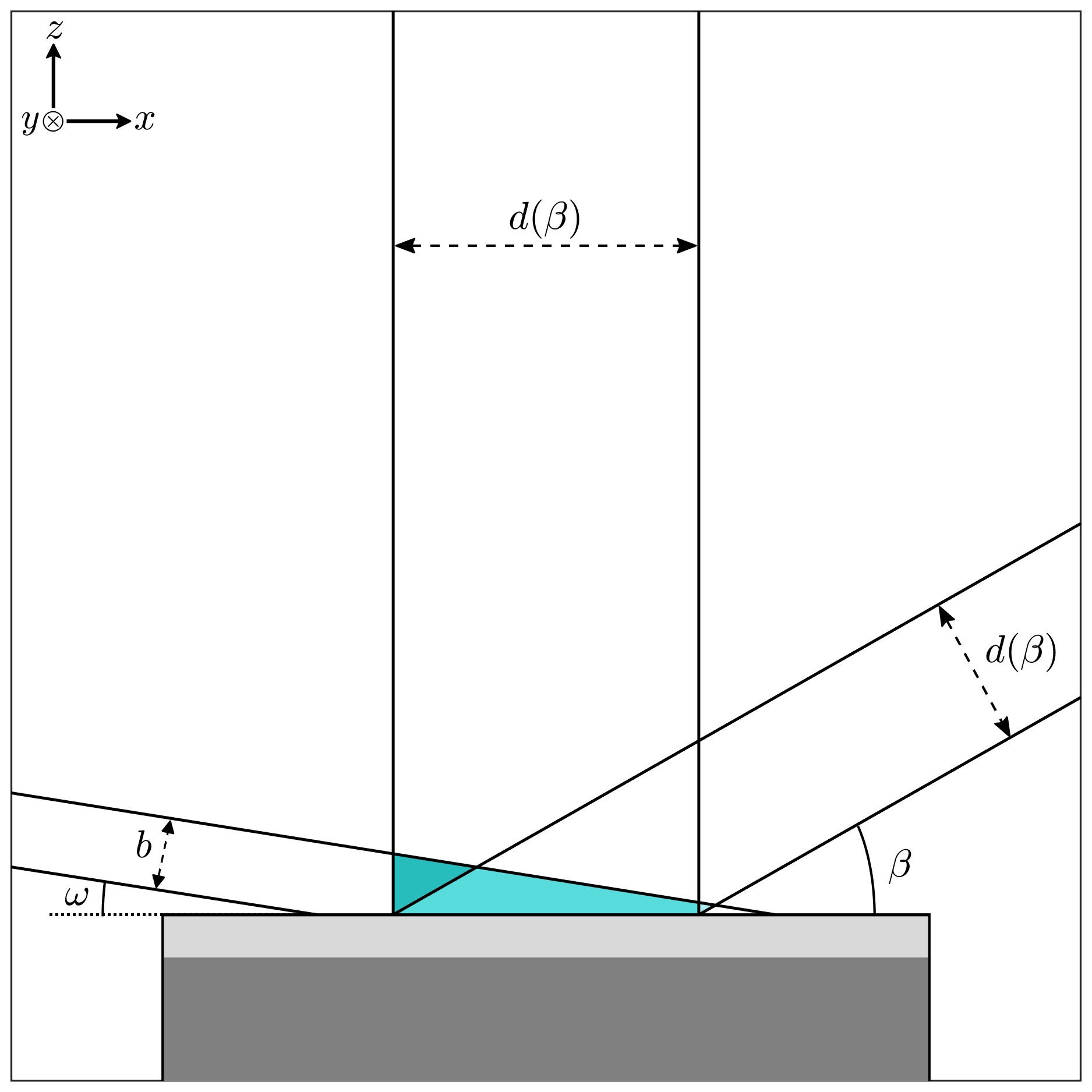}
	\caption{A schematic illustration of the grazing incidence setup involving a fixed incidence angle $\omega$, source slit $b$ and a variable detector angle $\beta$. The detector slit $d(\beta)$ can be adjusted to maintain constant sample area detection. The color highlighted region represents the $\beta$-dependent air scattering volume. }
	\label{fig:SetUp}
\end{figure}

In recent years, impressive breakthroughs have been made in obtaining the pair distribution functions from thin films utilizing synchrotron radiation in either transmission \cite{Jensen} or reflection mode using a grazing incidence (GI) geometry \cite{Shyam,Dippel, Roelsgaard} (see Fig. \ref{fig:SetUp} GI geometry schematics), and multiple   software packages, such as RAD \cite{https://doi.org/10.1107/S0021889889002104}, PDFGetX3 \cite{Juhas:nb5046}, and GudrunX \cite{Soper2011GudrunNAG}, exist to convert the measured intensity into a PDF. However, a GI approach applied in a lab setting with Cu, Mo or even Ag radiation remains a big challenge due to the limiting flux combined with a high substrate contribution and low diffraction signal. 

The early work on the theory and some experimental aspects of measuring pair distribution function from supported and free-standing films can be traced to Wagner, who emphasized using a free-standing film or a substrate with as low atomic number 

\noindent
as possible~\cite{10.1116/1.1315719}. Eguchi \emph{et al.} \cite{Eguchi} determined the pair distribution functions of 500 nm thick amorphous indium-zinc oxide films utilizing an x-ray tube with Mo-K$\alpha$ radiation. However, in order to obtain the isolated signal of the film they had to perform two additional scans; one of the substrate and another of the air. Not until these contributions were weighted and subtracted off from the total scattering intensity of the sample, were they were able to acquire the diffraction pattern of the film. While this procedure, whose success depends on accurate placement and alignment of the samples, represents the current state-of-the-art measurement protocol \cite{Dippel, Roelsgaard, Shyam}, it is clearly highly desirable to be able to extract the pair distribution function using a single measurement only.

Hence, in this work, we investigate the necessary measurement conditions and outline the procedures required to obtain accurate PDFs from sub-micron thin metallic glass films down to at least 80 nm thick films, with lab-based x-ray sources. Utilizing the recent advancements in x-ray optics and modern detectors, together with innovations in data analysis and sample design, the procedure becomes eminently straightforward. We demonstrate that it is possible to suppress, or even eliminate, the coherent substrate signal by utilizing a crystalline substrate. By orienting the substrate in a manner that avoids any Bragg reflections, the film signal can be isolated at the measurement step without requiring post-process substrate reduction techniques. Lastly, we examine the ultimate film thickness limit in which reliable amorphous PDFs can be determined in a lab environment before the scattered signal from the film is overshadowed by background noise. We have assessed existing practises and techniques on grazing incidence diffraction and developed new ones, which are collected and summarized in this work.

\section{Experimental details}
\subsection{Sample Growth}
The films were grown by DC magnetron sputtering onto a single-crystalline 0.5 mm thick and 20$\times$20 mm$^2$ A-cut Al$_2$O$_3$ (11$\bar{2}$0) substrate in an ultra-high-vacuum chamber with a base pressure of less than $3 \cdot10^{-10}$ Torr. To minimize the level of contaminants, argon of 6N purity was sent through a Nupure Omni 40 PF purifying filter in between the bottle and the sputter chamber. The substrates were annealed at 573 K for 30 min in ultra-high vacuum to reduce the amount of water and other impurities on their surface, and were subsequently cooled to room temperature before vanadium and zirconium were co-sputtered. Throughout this work we used  amorphous V$_{33}$Zr$_{67}$ of nominal thicknesses $t = \{324, 162, 81, 41, 10\}$ nm. To protect the samples from oxidation a thin $6$ nm layer of amorphous Al$_2$O$_3$ was deposited on top of the films. Rutherford backscattering spectrometry was used to verify the composition to within 1 atomic \% of the intended composition and x-ray reflectometry was used to determine the thickness of the samples. The samples were grown without the use of clamps for sample fixture, which ensures no visible areas of the bare substrate. Because the sample holder was rotated during growth, the sides of the substrates were also deposited.

\subsection{Sample model}
To help assess the validity of the experimentally obtained pair distribution functions, theoretical \emph{ab initio} density functional calculations of the pair distribution function, and its inverse Fourier transform to a theoretical structure factor, were computed. The first-principles calculations based on density functional theory~\cite{10.1103/physrev.140.a1133, 10.1103/physrev.136.b864}, as implemented in the Vienna Ab initio Simulation Package (VASP)~\cite{10.1103/physrevb.54.11169, 10.1016/0927-0256(96)00008-0, 10.1103/physrevb.47.558} along with the generalized gradient approximation (GGA) correlation-functional of Perdew, Burke and Ernzerhof,(PBE)~\cite{10.1103/physrevlett.77.3865} were used to construct amorphous candidate structures via the stochastic quenching procedure~\cite{10.1103/physrevb.82.024203}. All details on the model and convergence criteria can be found in the work of \cite{PhysRevB.106.104110}.

\subsection{Instrument description}
 All diffraction measurements were performed on a Malvern Panalytical third generation Empyrean instrument equipped with a 240 mm radius goniometer, using a long fine-focus Mo-$K\alpha$ source. The generator was set to 60 kV and 40 mA for all measurements. The instrument is equipped with an elliptically focusing multilayer mirror. The mirror illuminates a smaller sample area as compared to a  parallel mirror, thereby providing higher flux for thin films of small length along the beam direction. A divergence slit of 1/16 degrees was placed directly after the tube and a 0.02 rad Soller slit was placed after the mirror to collimate the beam and minimize illumination perpendicular to the scattering plane. A 1/4 degree anti-scatter slit was mounted after the Soller slit to reduce diffuse scattering from the mirror. A 20 mm beam mask was inserted on the incidence side to further restrict the width of the beam, and thereby also reduce air scattering. On the detector side, a parallel collimator of 0.28 degrees as found in the dCore system, was used together with 1Der detector operating in fixed static 1D mode. The collimator reduces air scattering as well as limiting the angular divergence of the experiment. A batch file was written in which the image of the detector was saved for every step taken in $2\theta$ (see Fig.\ref{fig:Air}(b)). The image files were processed with a specially written MATLAB code using the open source xml library~\cite{XRDMLread_cite}. The code for processing thin film diffraction patterns into pair distribution functions as well as the batch file can be found in Ref.~\cite{ThinRedPDF_cite}.
 
\section{Results}\label{sec:Theory}
\subsection{Instrument and sample considerations}\label{subsec:instr}

The contributions to the measured intensity $I_{\text{meas}}(2\theta)$, where $2\theta$ is the scattering angle (sum of the incidence angle $\omega$ and the detector angle $\beta$), is taken from Thijsse's~\cite{Thijsse} seminal paper but adapted for a film and substrate system and is given by:

\begin{equation}
\begin{aligned}
I_{\text{meas}}(2\theta)  =~ & I_{\text{f,coh}}(2\theta)   &+& I_{\text{f,com}}(2\theta)        + & \\
                             & I_{\text{s,coh}}(2\theta)   &+& I_{\text{s,com}}(2\theta)        + & \\
                             & I_{\text{air,coh}}(2\theta) &+& I_{\text{air,com}}(2\theta)      + & \\ 
                             & I_{\text{holder}}(2\theta)  &+& I_{\text{fluorescence}}(2\theta) + I_{\text{dark}},
\end{aligned}
\label{eq:I_meas}
\end{equation}

\noindent 
where $I_{\text{f,coh}}(2\theta) + I_{\text{f,com}}(2\theta)$ is the coherent and incoherent (Compton) scattered intensity from the film, $I_{\text{s,coh}}(2\theta) +I_{\text{s,com}}(2\theta)$ is the coherent and Compton scattering from the substrate, $I_{\text{air}}(2\theta)$ is the intensity from air scattering, $I_{\text{holder}}(2\theta)$ is any stray contribution from the instrument or sample holder,  $I_{\text{dark}}$ represents dark counts, and $I_{\text{fluorescence}}(2\theta)$ is the fluorescent intensity. We have neglected small angle scattering as well as multiple scattering as these are assumed to be small in thin films. These effects may be included using standard formulas and techniques. For the other contributions, it is important to note that due to the grazing incidence geometry, the illuminated area is fixed by the choice of incidence angle, source slit, mirror and beam divergence. Preferably, one would like to set an incidence angle which maximizes the thin film signal, while suppressing the influence of the substrate, sample holder and air. 

\subsection{Film and substrate fluorescence}\label{subsec:flour}
To minimize the influence of film and substrate fluorescence we used the 1Der detector from Malvern Panalytical capable of electronic photon energy filtering windows as narrow as 340 eV (Lynxeye XET from Bruker is another choice). Using modern electronic energy discrimination in this way also negates the need for using a Ni (for CuK$\alpha$) or Zr (for MoK$\alpha$) filters to further reduce the $\beta$ radiation. Furthermore, the discrimination replaces a secondary monochromator, which significantly improves the measured count rate. The energy filter also eliminates much of the Bremsstrahlung.

\subsection{Sample holder scattering}\label{subsec:Holder}
Since thin films are not always deposited on large wafers but quite often square-shaped substrates of either 10$\times$10 mm or 20$\times$20 mm size, it is important, given the small angles involved in grazing incidence, to minimize any spurious scattering from the instrument or sample holder. To this end, an in-house 3D printed holder was designed and printed, whose dimensions of the contact area with the supported sample are significantly smaller than the actual sample. The film is held in place by air suction, eliminating the need for adhesives or clamps, which further restricts any sample holder and signal contamination. The holder was manufactured using a \texttt{Form} 2 stereolithographic printer, using a rigid resin from the engineering resin family of materials available by \texttt{formlabs} (\href{https://formlabs.com/}{https://formlabs.com/eu/})
An illustration and/or computer-aided design files are available upon request. A beam mask and a Soller slit, combined with the focusing mirror and the divergence slit, kept the over-illuminated area of the sample, as well as the illuminated area of air, to a minimum. This approach avoids using any kinds of clamps that could partially shadow the sample surface and give rise to additional parasitic scattering.

\subsection{Air scattering}\label{subsec:air}
As noted by Warren~\cite{10.1107/s0021889870005654}, when measuring in a symmetric $\theta/2\theta$ mode, the air scattering contribution with the sample in place is half of the air scattering measured without the sample. This fact led to a standard procedure to eliminate the air scattering signal by separately measuring the air scattering without the sample and subtract off half of the air scattering. Unlike the symmetrical $\theta/2\theta$ scan, whereby the volume of air is kept approximately constant, the detectable volume of air varies throughout a grazing incidence measurement, making Warren's relation no longer valid. In fact, due to the absorption factor of the film, the air scattering contribution does not cancel when subtracting off the intensity from the bare substrate. In principle, one is therefore forced to perform three separate measurements under identical conditions and ensure that the alignment and placement of the samples are the same; one of the film/substrate system, one with only the substrate in place, and one accounting for the air contribution alone.

Fortunately, the volume of air above the sample is much lower when the grazing

\begin{figure}
	\centering
	\includegraphics[width= 0.65\columnwidth]{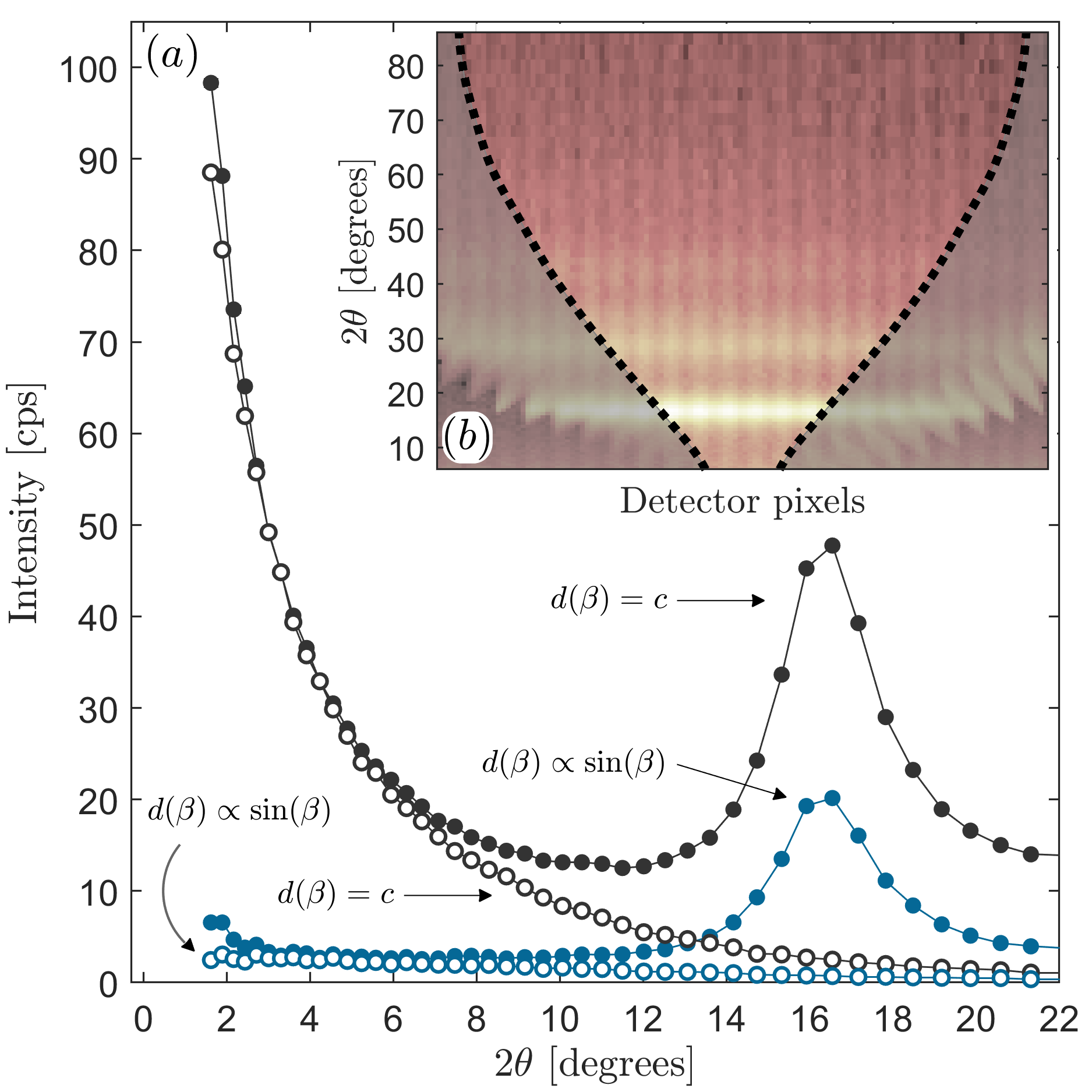}
	\caption{(a) Measured air (open circles) and sample (filled circles) intensity, with and without variable 1D pixel detector selection, where the number of averaged pixels are either chosen to follow $\sin{\beta}$ or to be constant $c$ throughout the scan. (b) The 1D detector image together with the outlined condition for constant detection footprint as the dotted line. Due to the use of the parallel collimator, intensity striping appears. }
	\label{fig:Air}
\end{figure}

\noindent
 angle is small, and is largest at low $2\theta$ values (see Fig.  \ref{fig:SetUp}). Under the assumption that the accepted (by the detector) scattered beam is parallel (nearly the case given the parallel plate collimator and the small divergence slit in front of the mirror), the angular dependence of the width of the diffracted beam is known, and was recently summarized by Rowles and coworkers~\cite{Rowles}. In brief, the visible region on the sample that a fixed detector opening observes during the angular range of a scan changes as $\propto 1/\sin{\beta}$, which implies that at very low angles, the detectable region becomes large, while at 90 degrees, that region constitutes approximately the detector width (8.89 mm in our case).

We exploited this fact by measuring the scattered intensity with the 1DER dector in fixed static 1D mode, and selected a post-measurement selection window on the 1D image such that a constant region throughout the entirety of the scan was seen by the detector, see the non-shaded region in Fig.~\ref{fig:Air}(b). In other words, by allowing the number of pixels to change proportionally to $\sin{\beta}$, the detected sample area is kept constant throughout the scan, rendering the air volume approximately constant. By virtue of the low grazing incidence angle, the volume of air, and hence its scattering contribution, will therefore be significantly suppressed. However, as the film becomes thinner (and the intensity from the film decreases), this effect may longer be negligible, and other options, such as evacuating the air or replacing it with helium, can be considered. Fig.~\ref{fig:Air} shows the scattering from the sample including air for two cases. On the one hand when the detector opening remained constant and on the other, when the detector opening was varied as described above. Also shown in the figure is the air scattering without the sample (open symbols). In the inset, a composite image of the individual line scans is shown with the part of the detector which is accepted highlighted by the interior between the dashed lines.  By employing the variable pixel window, we are able to lower the air scattering detection by a factor of at least 90. The apparent decrease in intensity of the amorphous peak by a factor 2.5 is, as pointed out by Rowles and co-workers, a consequence of a geometrical aberration, which is implicitly accounted for in the variable detection case. Correcting the data with the constant detection opening for this aberration, one finds that the intensity is actually lower than the one using the variable pixel selection procedure. The advantage of this procedure is that it essentially eliminates a third separate scan of the air scattering if the film is sufficiently thick. At the lowest angles we also see some off-specular scattering from roughness of the film, which limits how low in angle one can measure.

\subsection{Influence of the substrate}\label{subsec:substr}
Due to the penetration of the x-ray beam into the substrate, the feasibility of a single-shot experiment in reflection is predominantly decided on the material properties and structure of the substrate, and the thickness of the film. For instance, the use of an amorphous substrate will impart its own coherent scattering term, $I_{\text{s,coh}}(2\theta)$ (as well as  $I_{\text{s,com}}(2\theta)$), regardless of measured scattering wavevector $Q=|Q|=4\pi/(\lambda\sin(\theta))$ as its scattering conditions only depend on the magnitude of the scattering vector. Thus, one is forced to subtract the amorphous substrate signal which requires a second substrate scan and precise knowledge about the attenuation through the layers of the film. On the other hand, if one considers a single crystalline substrate, the discreteness of the reciprocal lattice in conjunction with the asymmetric grazing incidence geometry allow for momentum transfer values, $(Q_x, Q_y, Q_z)$, that trace a path in reciprocal space that does not intercept lattice points on the surface of the Ewald sphere. This should eliminate coherent Bragg scattering contributions from the substrate, and therefore the need of substrate subtraction altogether. However, it should be noted that if the path ends up close of fulfilling the Bragg conditions, a small contribution might still add to the detected intensity, owing to scattering associated with lattice vibrations or defects in the crystal. It is possible to bypass the crystal reflections by utilizing a crystal with a low-symmetry surface cut, as the additional restriction in symmetry further constrain the number of available scattering planes in grazing incidence configuration. Tilting or rotating the sample is also a method of tweaking the reciprocal path with an offset in $Q_y$. Ideally one would want an obliquely cut crystal such that it does not give rise to any Bragg peaks at the chosen wavelength. We assessed the ability of A-cut (11$\bar{2}$0) sapphire to fill this role in this work, which turned out to be successful, as will be discussed. One should note that the choice of a crystalline substrate often does not hamper the growth of amorphous compounds 

\begin{figure}
	\centering
	\includegraphics[width= 0.95\columnwidth]{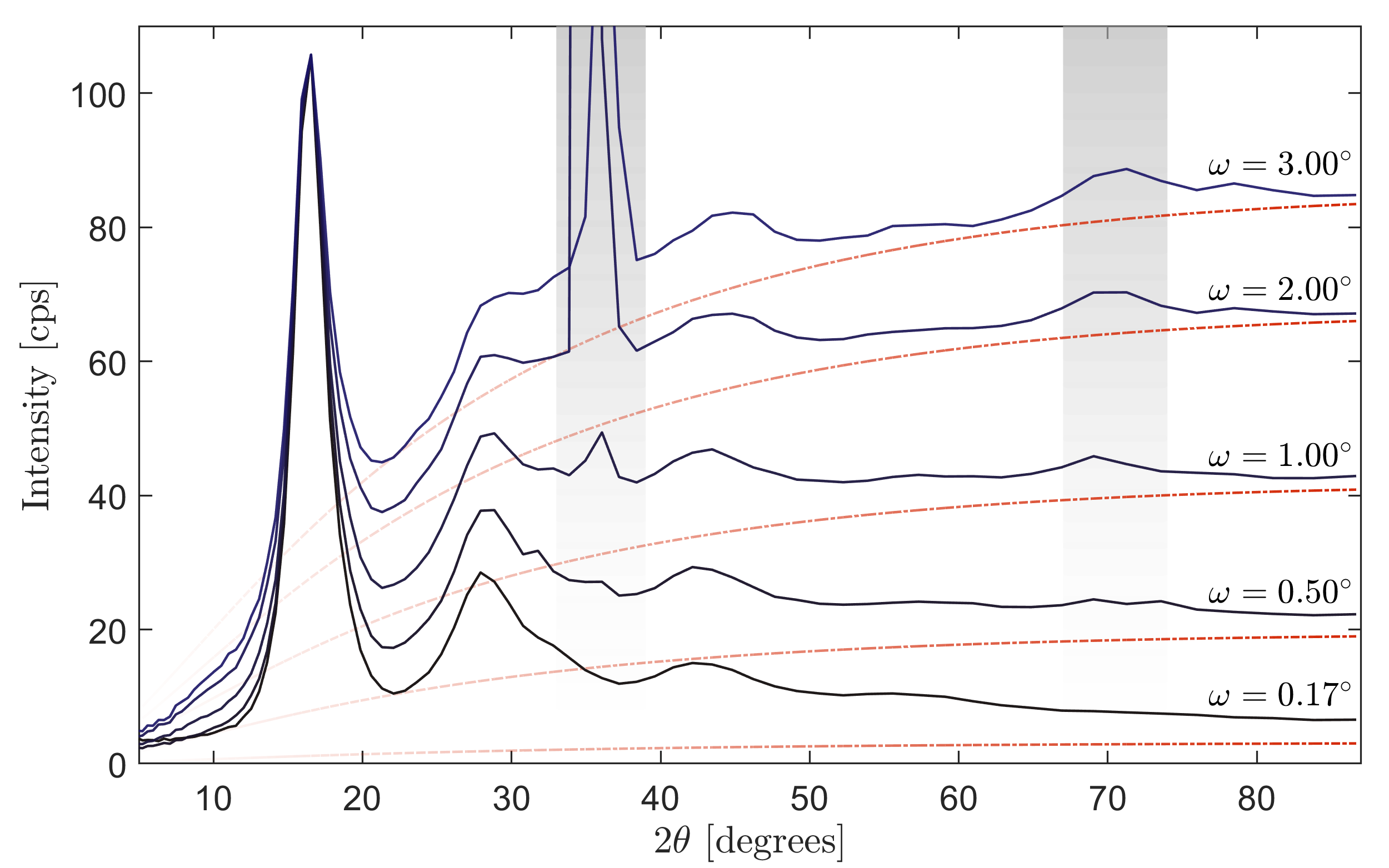}
	\caption{Series of GIXRD scans with different choice of incidence angle, normalized to the first amorphous peak for clarity. The red dashed lines correspond to the Compton intensity of the substrate. The gray regions highlight the presence of Bragg peaks at high angles which are eliminated at lower $\omega$ angles.}
	\label{fig:I_omega}
\end{figure}

\noindent
owing to the possibility to grow a thin wetting or seed layer between the film of interest and the substrate~\cite{10.1016/j.tsf.2010.07.084}. Such thin wetting or seed layers are used to eliminate possible crystal nucleation sites and promote amorphous growth of the glassy compound.

Although the aforementioned approach is designed to evade the Bragg scattering intensity of the substrate, the incoherent Compton scattering from the substrate is not removed by this procedure. Fig.~\ref{fig:I_omega} illustrates the issue for a 325 nm film of V$_{33}$Zr$_{67}$ on 0.5 mm A-cut (11$\bar{2}$0) sapphire. The detected intensity is shown as a function of the scattering angle for a series of incident angles $\omega$. At $\omega=3.00$ degrees we note, firstly, the presence of two Bragg peaks, and secondly, the rising background intensity, which follows the expected Compton scattering profile of sapphire. As we lower the incidence angle, both the Compton intensity and the Bragg intensity of the substrate goes down but at different rates. therefore, the Compton scattering is a good gauge of the amount of electric field intensity present in the substrate and the potential for Bragg scattering. This also highlights how the Bragg condition can be avoided by fine-tuning the incidence angle (in addition to the azimuthal and tilt angles of the sample). At the angle of $\omega=0.17$ degrees, the intensity shows no sign of Bragg scattering, and the Compton scattering is manageable for this thick film. Given the good energy filtering properties of modern detectors, we assessed the possibly of filtering out the remaining Compton scattering from the film and substrate.  Due to its characteristic energy shift (see Appendix and Eq.~\ref{eq:Compton_shift}), choosing a narrow energy window around the elastic line of (17.15 keV--18.00 keV), the Compton signal could be suppressed above scattering vectors of about 5 \AA$^{-1}$ . Below this, the Compton remains partly unfiltered (see Appendix~\ref{Appendix:SiO2}).

The advantage of filtering out the Compton scattering is that for thinner films, the high angle signal from the film is drowned out by the Compton scattering, such that even if compensated for by the known theoretical signal from the substrate, the fluctuations associated with the Poisson noise would always be present. Furthermore, owing to optical effects near the critical angle, it is not completely straightforward to know beforehand the exact ratio of substrate to film Compton scattering.

\subsection{Absorption}\label{subsec:abs}
The angle-dependent absorption correction for grazing incidence geometry was summarized by Rowles and references therein~\cite{Rowles}, and is here written for a film/substrate system in terms of number of formula units of the compound that are in the effective scattering volume adapted from Ref.~\cite{PhysRevB.44.498}:
\begin{equation}
A_n(\omega, \beta) = N_n\frac{b\times h}{\sin{\omega}}e^{-\sum \limits_{i=1}^{n-1} \mu_i t_i\left(\frac{1}{\sin\omega} + \frac{1}{\sin\beta}\right)}\int \limits_0^{t_n}e^{-\mu_n t\left(\frac{1}{\sin\omega} + \frac{1}{\sin\beta}\right)} dt,
   \label{eq:Absorb_common}
\end{equation}

\noindent
where $N_n$ is the number density of formula units of the compound in layer $n$, $\mu_i$ is the linear attenuation of layer $i$ in the film/substrate stack and $t_i$ is the thickness of layer $i$. The $1/\sin{\omega}$ prefactor accounts for the projection of the incident beam width $b$ onto the sample and $h$ is the size of the beam in the $y$ direction. The effective $b$ is modified both by the divergence of the source and the convergence of the mirror but, at low enough incidence angles, overillumination typically occurs. Eq.~\ref{eq:Absorb_common} expresses the number of formula units that are present in the effective scattering volume and the scattering from which is detected at angle $\beta$. The expression does not include the effects of roughness and x-ray optical changes near the critical edge.

\begin{figure}
\centering
\includegraphics[width= 0.75\columnwidth]{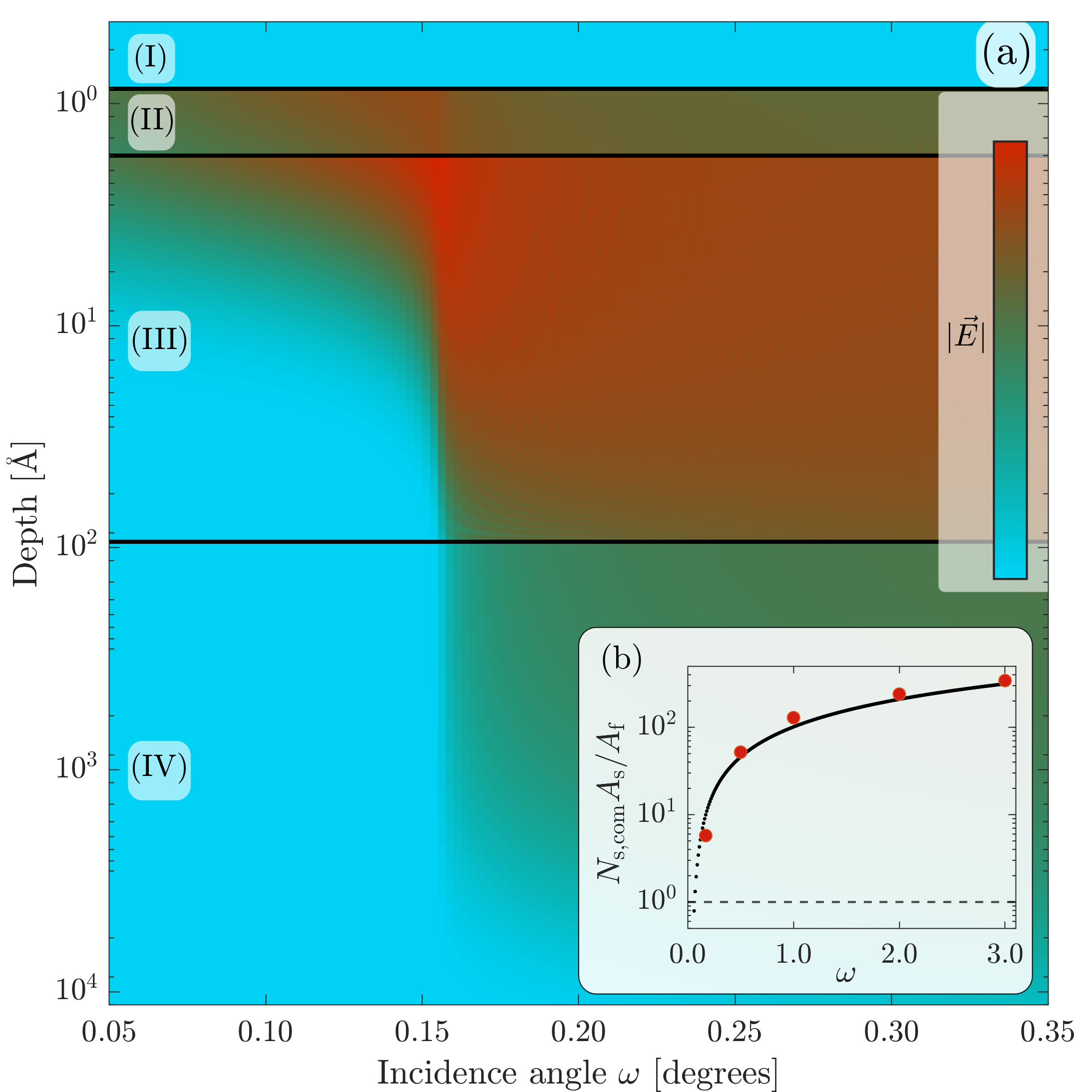}
\caption{Figure (a) illustrates the electric field intensity in the sample stack. Region $($I$)$ represent the air layer above the sample, $($II$)$ the protective Al$_2$O$_3$ layer, $($III$)$ the amorphous V$_{33}$Zr$_{67}$ layer, and $($IV$)$ is the substrate layer. The inset (b) shows the predicted absorption ratio $A_{\text{s}}/A_{\text{f}}$ between the substrate and film at different incidence $\omega$ angles, compared with the corresponding measured and fitted Compton profile prefactor $N_{\text{s,com}}A_{\text{s}}/A_{\text{f}}$ seen in Figure~\ref{fig:I_omega}, shown as red solid circles.}
\label{fig:depth} 
\end{figure}

To assess these effects we have calculated the absorption coefficient and its angular dependence using the methodology proposed by De Boer~\cite{PhysRevB.44.498, https://doi.org/10.1002/xrs.1300180309}, which takes reflectivity, multiple scattering, roughness at all interfaces, and absorption into account. Fig.~\ref{fig:depth} shows the electric field intensity as a function of depth and incidence angle for the film stack used in this work. We see the presence of the critical edge at around 0.14 degrees and of the well-known evanescent wave solution at shallower angles. As a comparison, the inset shows the calculated absorption ratio between the substrate $A_{\text{s}}$ and film $A_{\text{f}}$  using Eq.~\ref{eq:Absorb_common} as the black dotted line. The absorption factors were computed at different incidence angles and evaluated at $\beta = 90$ degrees. Together with the absorption ratio is the experimentally measured Compton profile prefactor $N_{\text{s,com}}A_{\text{s}}/A_{\text{f}}$, determined by fitting the measured intensity of Fig.~\ref{fig:I_omega} to the Compton profile in Eq.~\ref{eq:I_meas}. We see here for the 324 nm thick sample that the Compton intensity of the substrate is well-described by the amount of absorbed intensity in the substrate, and any deviating residual effects can assumed to be contained in $N_{\text{s,com}}$, which should be of the order of unity (see e.g. Ref.~\cite{Thijsse}). Hence, it is possible to predict the ratio of substrate Compton intensity to subtract in the data reduction process if the absorption factor in the substrate is known. However, the approximation in Eq.~\ref{eq:Absorb_common} may no longer be valid for ultra-thin films, especially in gracing incidence geometry close to the critical edge, where optical enhancement effects need to be treated in the absorption correction.

To be able to simulate the electric field strength with reasonable accuracy, one needs to know the composition, thickness and absorption of the material in each layer. Most of this information can be obtained from a fit for the x-ray reflectivity profile of the sample, such as implemented by the GenX program~\cite{bjorck_genx_2007}.  Once a reliable density profile is obtained, De Boer's formalism can be used to calculate the absorption in the film, substrate, and if present, the absorption in the protective top layer as well. These angle-dependent quantities are then used in the data reduction procedure to obtain the pair distribution function, along with the determined proportion of Compton scattering between the film and the substrate. 

\subsection{Choosing the incidence angle}

\begin{figure}
	\centering
	\includegraphics[width= 0.6\columnwidth]{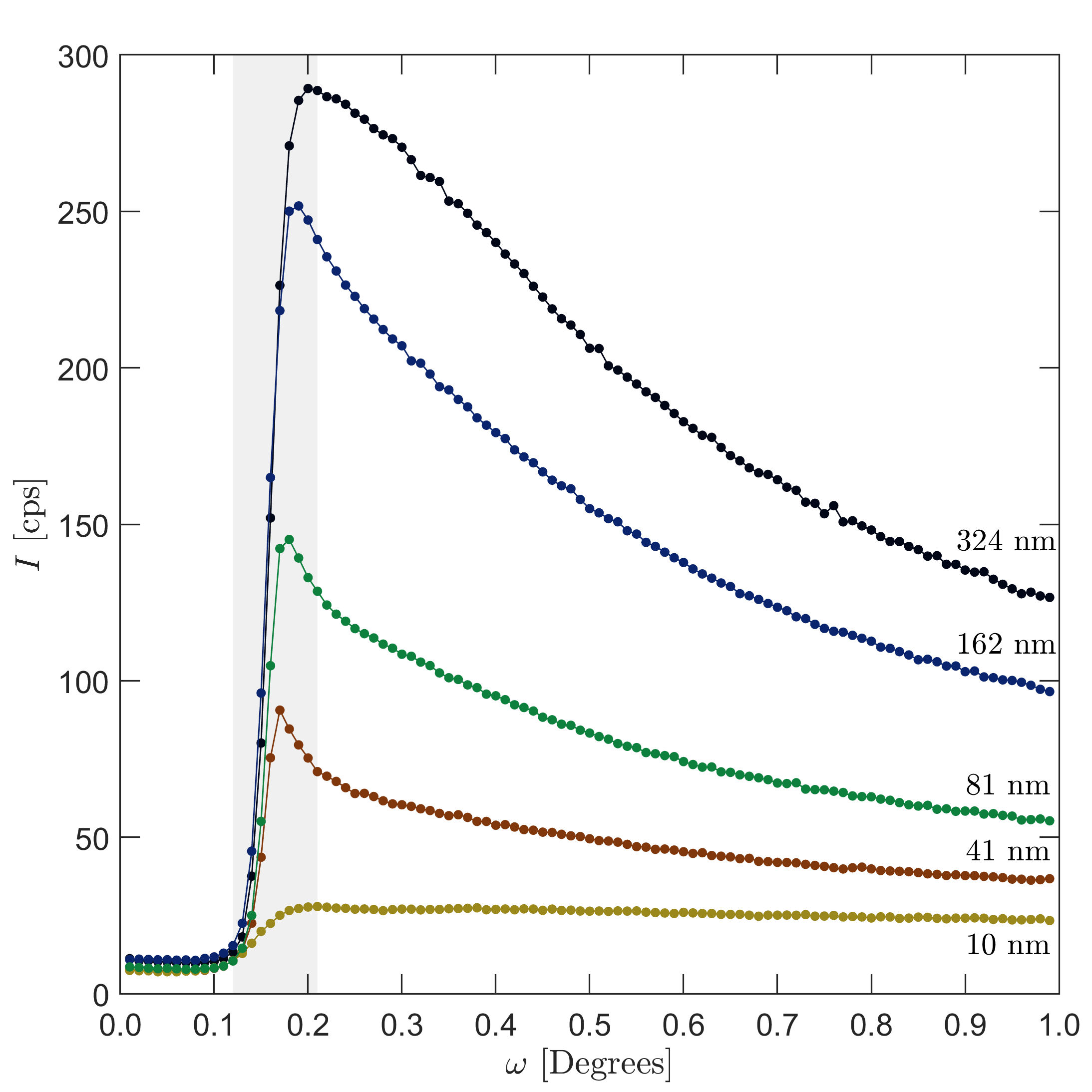}
	\caption{Low-angle $\omega$ scans with the detector angle fixed at the amorphous first peak (see Figure~\ref{fig:I_coh}) of V$_{33}$Zr$_{67}$ metallic glass, with a thickness of 324, 162, 81, 41, and 10 nm respectively. The flat region $\omega < 0.1$ degrees corresponds to conditions below the critical edge, and the intensity decrease seen in the region above above $\omega > 0.2$ degrees is due to overillumination and Compton scattering from the substrate.}
	\label{fig:I_H}
\end{figure}

From Fig. ~\ref{fig:depth} we reached the intuitive conclusion that one should choose an angle close to the critical edge. To test this experimentally, we measured a rocking curve at a fixed scattering angle corresponding to the first maxima in Fig. ~\ref{fig:I_omega}. At this angle, only weak Compton scattering from the substrate is expected. Fig.~\ref{fig:I_H} shows the results for different thickness of the V$_{33}$Zr$_{67}$ layer. We see that the intensity from the first maximum peaks close to the critical edge, to then further decay at larger angles. This decay is due to the 1/$\sin{\omega}$ factor in Eq.~\ref{eq:Absorb_common}. The maximum intensity coincides with the largest electric field strength build-up in the film, as seen in Fig.~\ref{fig:depth}. We note that the intensity ratio between the maxima and the intensity at 1 degree goes down as the film becomes thinner, since the Compton scattering from the substrate starts to dominate. In principle, it is the ratio of the film scattering to the substrate scattering that should be maximized, not only the film scattering. We found by trial and error that choosing the incidence angle slightly lower than the angle indicated by the maximum, is a good choice.

\subsection{Reduction of diffraction pattern to pair distribution function}\label{subsec:diff_to_PDF}

We are now in the position to combine all these considerations, i.e., utilizing a crystalline substrate, tuning the detector opening to a constant detector footprint, minimizing scattering from sample holder and fluorescence, and carefully choosing the incidence angle. What remains are the standard corrections for polarization, depth absorption profiles and known Compton contributions to equate the measured intensity to the elastic coherent scattering of the metallic glass film.

After subtracting dark counts, suppressing fluorescence, air scattering, and coherent scattering from the substrate, one obtains the relation for the coherent scattering of the film $I_{\text{f,coh}}$ as,
\begin{equation}
	I_{f,\text{coh}} =  N_0\frac{I_{\text{meas}}}{PA_{\text{f}}} - I_{\text{f,com}} - N_{\text{s,com}}\frac{A_{\text{s}}}{A_{\text{f}}}I_{\text{s,com}} 
	\label{eq:I_meas}
\end{equation}
\noindent
where $P$ is the polarization factor (Eq.~\ref{eq:Polarization}), and $A_{\text{f}}$ and $A_{\text{s}}$ are the absorption factors in the film, and substrate, respectively (Eq. \ref{eq:Absorb_common}). $N_0$ is an instrument-dependent normalization constant and $N_{\text{s,com}}$ corresponds to the weight of the Compton contribution of the substrate while $I_{\text{f,com}}$ and $I_{\text{s,com}}$ are the known, tabulated, film and substrate Compton profiles (see Eq.~\ref{eq:Compton}, \ref{eq:Compton_weight}, and \ref{eq:Compton_shift} for their parameterized form). The unknown quantities are the normalization constant $N_0$ and the Compton weight $N_{\text{s,com}}$. 
\begin{figure}
\centering
        \includegraphics[width= \columnwidth]{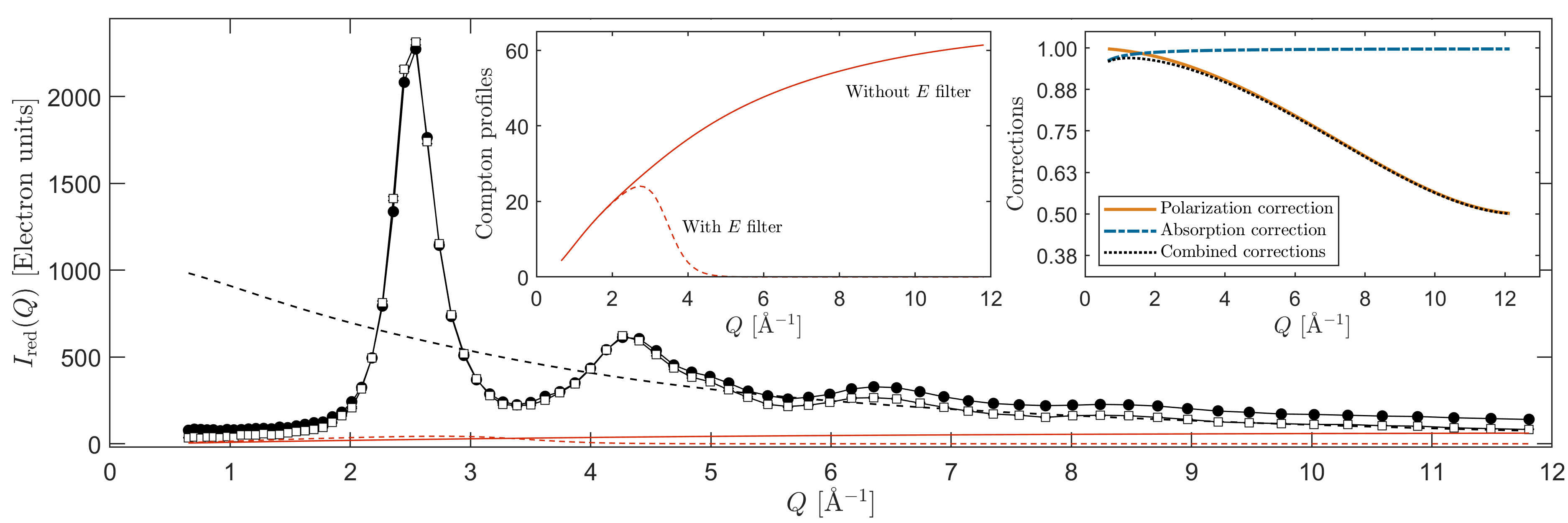}
        \caption{The measured normalized scattering intensity in electron units ($I_{\text{red}} = N_0I_{\text{meas}}/(PA_{\text{f}})$), with (open circles) and without (solid circles) energy filtering. The corresponding Compton profiles are shown in the left inset and applied correction factors in the right inset. The black dashed line is $\langle f^2 \rangle$, the red dashed dotted line correspond to the Compton intensity from the film and the solid red line represent the Compton intensity of the substrate.}
        \label{fig:I_coh}
\end{figure}

\noindent
As mentioned above, $A_{n}$ can either be calculated using the approximate expression Eq.~\ref{eq:Absorb_common} or by using the electric field intensity calculation to include x-ray optical effects. 

The prefactor, $N_0$, can be determined once for a given instrument configuration but can be also empirically found with the aid of known features of the scattering properties of amorphous materials. For instance, Wagner \cite{10.1116/1.1315719} developed a method exploiting the decrease in the atomic pair correlations attributing that the tail of the scan above a certain $Q$ value is bound to oscillate weakly around its squared form factor value, as also displayed in Fig.~\ref{fig:I_coh}. With this, one can determine the value of $N_0$, using Eq.~\ref{eq:Normalization_f_tail}. An alternative to this method is to make use of sum rules, such as the one shown in Eq.~\ref{eq:sum_rule_Q}, and the fact that the slope around the low $r$ region in the reduced pair distribution function $G(r)$ is expected to be linear, as shown in Eq.~\ref{eq:sum_rule_R}. However, we would like to point out that the sum rule of Eq.~\ref{eq:sum_rule_Q} is prone to enhance statistical errors for large $Q$-values and is unfortunately a slowly converging quantity with respect to $Q$. 

Instead, we used the slope of $G(r)$ (as also recommended by Thijsse~\cite{Thijsse}) at low $r$ as the leading figure of merit, as any false $N_0$ would yield nonphysical deviations from the ideally straight slope in this region. Furthermore, this feature also enables the coefficient $N_{\text{s,com}}$ to be determined, as any unknown stray background functions embedded in the structure factor $F(Q)$, such as a wrong Compton profile, introduces erroneous $Q$-dependent contributions into $F(Q)$, which give rise to real-space oscillations in this low $r$ region. In effect, by minimizing these deviations from a straight line, one is able to pinpoint the optimum value of $N_0$ and $N_{\text{s,com}}$ for the most physically sound pair distribution function. We wrote a predictor tool~\cite{XrayDepthProfiler_cite} as part of the data reduction that not only estimates $N_{\text{s,com}}$, but also aids in guiding the data reduction process towards a physically reasonable set of bound parameters.

Once the film $I_{\text{coh}}$ has been obtained, the structure factor $F(Q)$ can be computed with standard methods via the Zernike-Prins formula~\cite{Zernike1927} following:
\begin{equation}
F(Q) = \frac{I_{\text{f,coh}} - \langle f^2 \rangle }{\langle f \rangle^2},
\label{eq:F}
\end{equation}
\noindent
where $f(Q)$ is the form factor of the metallic glass (see appendix Eq.~\ref{eq:formFactor},\ref{eq:nthformFactor}, \ref{eq:formFactor_av_sq} and~\ref{eq:formFactor_sq_av}). From the structure factor, one can via a spherically symmetric Fourier transform compute the reduced pair distribution function $G(r)$ via:
\begin{equation}
G(r) = \frac{2}{\pi} \int \limits_0^{Q_\text{max}} QF\sin(Qr) dQ.
\label{eq:redG}
\end{equation}
Bearing in mind that truncation errors due to a finite $Q_\text{max}$ might show up in $G(r)$, one can reduce these by either dampening the tail of $F(Q)$ \cite{Mozzi1969TheSO}, or employ a Lanczos filter, (see Eq. \ref{eq:Landczos})~\cite{Thijsse}, to smoothly convolute the real-space function $G(r)$ and remove these artificial oscillations. 

With the reduced pair distribution function, $G(r)$ can be converted into the true pair distribution function $g(r)$ using:
\begin{equation}
g(r) =  \frac{G}{4\pi\rho_0 r} + 1.
\label{eq:g}
\end{equation}
All auxiliary functions, equations and constants used are summarized in the Appendix~\ref{sec:appendix} for the convenience of the reader.

Taking the $324$ nm thick sample as a starting point, and using the method of optimizing the low $r$ region of $g(r)$, the polarization and absorption corrected measured scattering intensity, i.e. $I_{\text{red}} = N_0I_{\text{meas}}/(PA_{\text{f}})$, is displayed in Fig.~\ref{fig:I_coh}. Two scans are portrayed, one with a narrow energy filter (17.15 keV--18.00 keV), shown here as hollow black circles, and one accepting a wider range of photon energies (14.00 keV--18.00 keV), displayed as the solid black circles. In the inset figures we show their respective total Compton profiles along with the correction factors, $P$ and $A_n$. In the inset showing the Compton profiles we can see that the two intensities agree well with one another up to a wavevector value of $\sim 4.5~$\AA$^{-1}$, at which point the scan utilizing a broader energy filter starts to trail above the one employing the strict energy discrimination due to the inclusion of unfiltered Compton scattering. Their difference is a testament to the effective elimination of Compton intensity when a narrow energy discrimination window is utilized.  Though, correctly accounting for the Compton intensity is still a necessary step, and depending whether the energy filtration is employed, either a complete Compton profile, or a profile that has been multiplied with an error function (like the one in the inset of Fig.~\ref{fig:I_coh}), have to be used. 

\begin{figure}
\centering
        \includegraphics[width= \columnwidth]{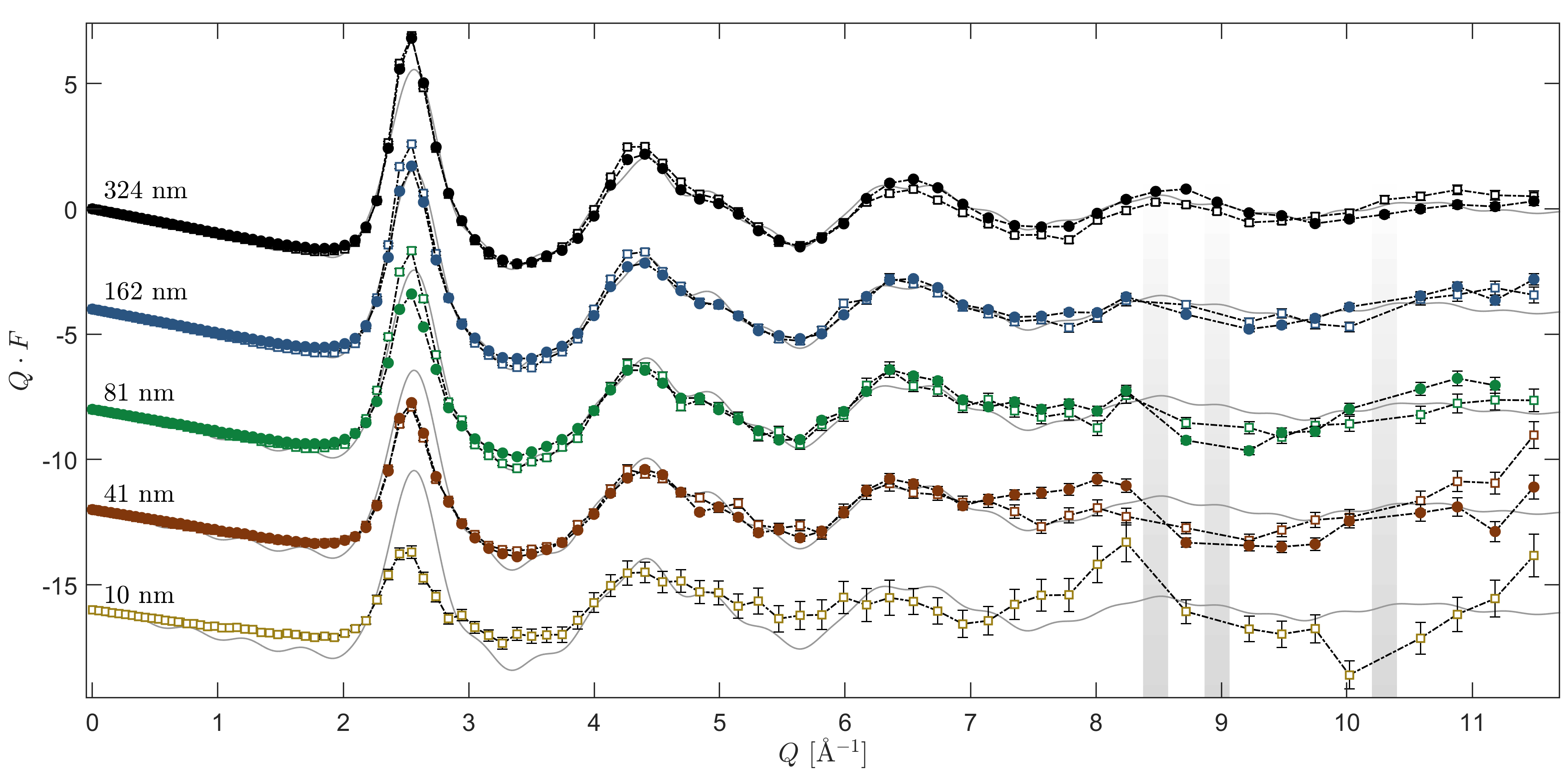}
        \caption{The structure factors $F(Q)$ from amorphous V$_{33}$Zr$_{67}$ of nominal thickness $t = \{324, 162, 81, 41, 10\}$ nm. The solid circles are measured using a wide energy discriminator window while the hollow circles are measured with a narrow energy window. The gray regions indicate the presence of minor stray Bragg peaks. The gray solid lines correspond to computed theoretical structure factor via \emph{ab initio} density functional theory. The structure factors have been shifted vertically for clarity.}
        \label{fig:F}
\end{figure}

To illustrate more clearly the impact of decreasing sample thickness on the feasibility of accurately measuring the diffraction patterns of thin metallic glass films, the determined structure factors $F(Q)$, derived from scans utilizing both a narrow energy filter (open circles) and wide energy window (solid circles), of the five sample sets are displayed in Fig.~\ref{fig:F}, along with the calculated theoretical prediction \emph{ab initio} stochastic quenching method in solid gray line. At this stage, we can see excellent agreement between the experimental results and theory, especially for the $324$ nm thick sample. By using the theoretical predictions as a figure of merit, we can see that we have good 

\noindent
agreement and consistency in a region $Q < 6 $~\AA$^{-1}$ for virtually all sample thicknesses except for the 10 nm one. Though, what is also apparent, is that the data $Q > 7 $~\AA$^{-1}$ becomes more and more noisy, and its severity scales with the diminishing thickness of the samples. Minor substrate Bragg peaks make their appearance in this region that we were unable to completely eliminate. This might be remedied by another  surface cut or choice of substrate. Data points corresponding to the crystalline peaks have been removed as indicated by the gray lines. Even if the Compton profile is accurately subtracted for the unfiltered scan, the fluctuations associated with the shot noise can never be removed. Therefore it is more desirable to filter out the Compton scattering if possible.

\begin{figure}
	\centering
	\includegraphics[width= \columnwidth]{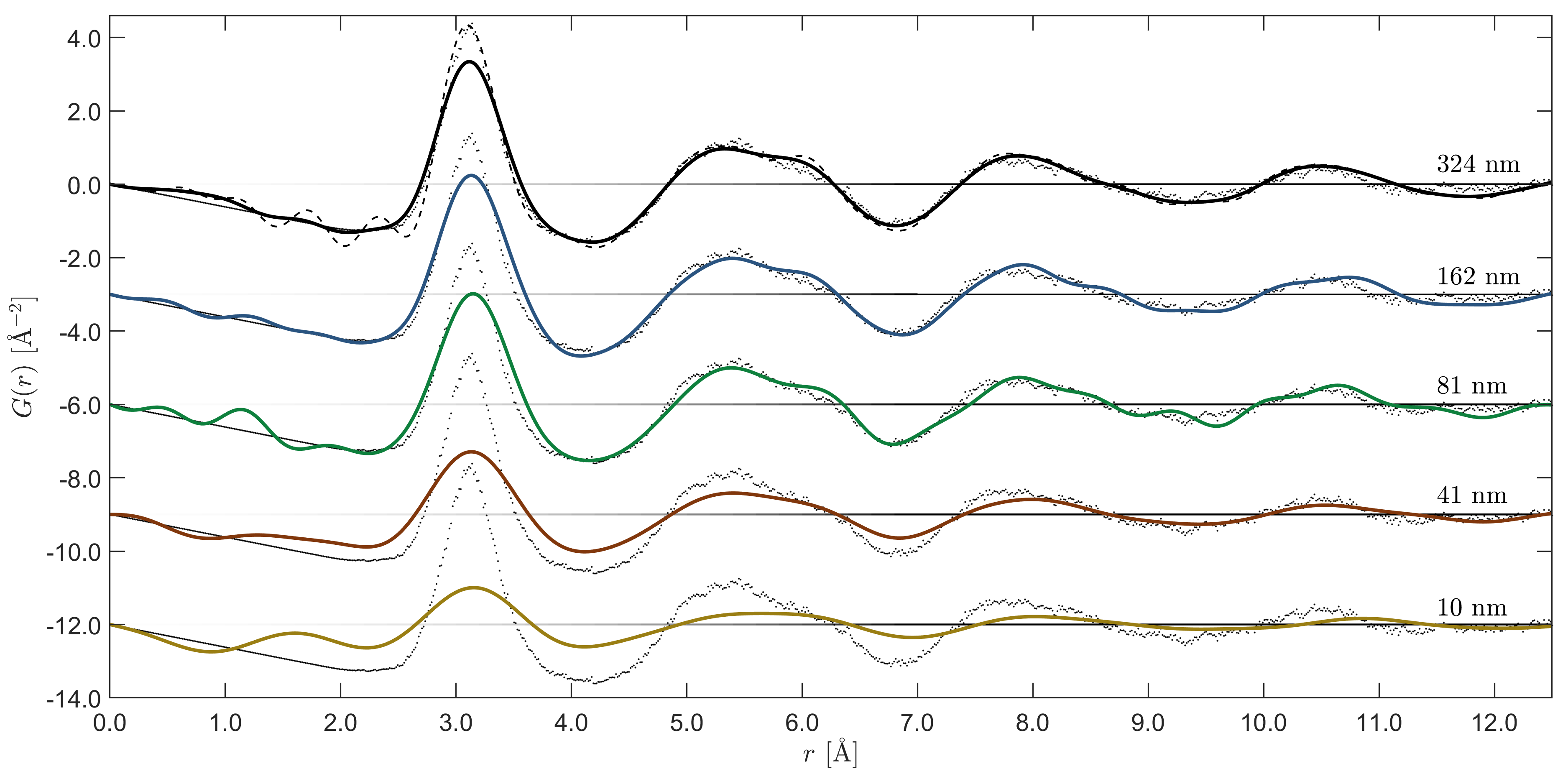}
	\caption{The reduced pair distribution functions $G(r)$ from amorphous V$_{33}$Zr$_{67}$ of nominal thickness $t = \{324, 162, 81, 41, 10\}$ nm, determined from the set of structure factors measured with a narrow energy window and transformed up to $Q_\text{max} = \{11.5, 11.5, 11.5, 8, 7\}$~\AA$^{-1}$ respectively. The gray dots are the theoretical predictions. The reduced pair distribution functions have been shifted for clarity.}
	\label{fig:G}
\end{figure}

We can however draw some conclusions by virtue of the consistency for $Q < 6 $~\AA$^{-1}$ of the structure factors. The significant reduction in reciprocal scattering correlations with increasing $Q$, i.e. the diminishing height of the oscillations, can be used as a guideline to assess at what value of $Q$ we reach the point of diminishing returns in terms of the information content of the scan. Bearing in mind that the structure factors have been multiplied by $Q$ which significantly amplifies these minuscule changes, the relative height of the last significant peak ($Q \approx 8.7$~\AA$^{-1}$) for the 324 nm thick sample has reached such an amplitude, that it barely contains any structural information beyond what is supplied by the uncorrelated form factor of the material. This means that for this system the choice of $Q_\text{max}$ beyond this point only marginally improves the informational content reflected in the PDF. However, this cut-off is highly system-dependent and can vary between the choice of amorphous materials (such as in Ref.~\cite{Eguchi, Dippel, Roelsgaard, Shyam}). For instance, amorphous SiO$_2$, measured by Warren~\cite{Mozzi1969TheSO}, has clear and distinct reciprocal correlations for $Q > 20 $~\AA$^{-1}$, and only at the very largest wavevector values does he find sufficient convergence. We compare Warren's measured structure factor with our measurement of bulk amorphous SiO$_2$ in GI geometry in the Appendix to illustrate the validity of the outlined approach.

Finally, the reduced pair distribution functions, obtained by Fourier transforming the energy-filtered structure factors via Eq.~\ref{eq:redG}, are displayed in Fig.~\ref{fig:G} along with our theoretical prediction (gray dots). We have chosen to present the data with a Lanczos filter to suppress termination ripples, with the drawback being of broadening all features~\cite{Thijsse}.  For reference, the black dashed curve shown together with the reduced pair distribution function of the 324 nm thick sample show how the data would look like without the Lanczos method. Here, one can see that for thicknesses above 81 nm, we achieve close agreement with theory. For the 41 nm case, we observe almost the same level of agreement as for the thicker sample measurements, while the thinnest 10 nm fails to deliver a reliable representation of the pair distribution function.

\section{Conclusion and outlook}
We have shown that the pair distribution function of thin metallic glasses can be measured down to at least 80 nm by considering systematic improvements of several areas regarding how the pair distribution functions are measured in the grazing incidence geometry. Owing to the x-ray scattering of materials being proportional to the square of the atomic number and the studied films having an average atomic number of about 34, highlights the applicability of the method to a wide range of materials. Further improvements can be gained by a judicious choice of substrate, higher quantum efficiency and a larger detector window. Further small gains may be found by using a parallel plate collimator with a larger angular divergence. These results are useful for structural analysis of thin films, which may have correlations different than those found in the bulk. The method can be used as pre-screening for further studies at synchrotron facilities, whereby the pair distribution function can be measured much faster and with higher real-space resolution. These findings clear the path for the possibility of utilizing standard modern diffraction equipment to determine thin film pair distribution functions.

\appendix
\section{Collection of expressions and formulas used in the data reduction}
\label{sec:appendix}

The incident angle $\omega$ and detector angle $\beta$ yield the scattering angle, $2\theta$, which can be expressed in terms of its corresponding wave vector modulus:
\begin{equation}
\label{eq:q}
Q \equiv |\overrightarrow{Q}| = 4\pi \sin(2\theta/2)/\lambda,
\end{equation}
\noindent
where $\lambda$ is the wavelength of the x-ray beam. 

The normalization procedure, developed by Norman \cite{10.1107/s0365110x57001085,10.1116/1.1315719}, adapted to film and substrate as:
\begin{equation}
\label{eq:Normalization_f_tail}
N_0 = \frac{ \int\limits_{Q_\text{min}}^{Q_\text{max}} \left[ \langle f^2 \rangle + I_{\text{f,com}} + N_{\text{s,com}}\frac{A_{\text{s}}}{A_{\text{f}}}I_{\text{s,com}}\right]} {  \int\limits_{Q_\text{min}}^{Q_\text{max}}  I_{\text{meas}}/(PA_{\text{f}}) },
\end{equation}
\noindent
whereby $Q_\text{min}$ corresponds to the lower bound where the modulations are weakly oscillating around its squared form factor value, and $Q_{\text{max}}$ represents  the upper bound considered in the Fourier transform of Eq. \ref{eq:redG}.

The structure factor sum rule, from Thijsse \cite{Thijsse} and Wagner \cite{WAGNER19781} is:
\begin{equation}
\label{eq:sum_rule_Q}
\int\limits_0^{Q_{\text{max}}} Q^2F(Q)dQ = -2\pi^2\rho.
\end{equation}
\noindent
in which $\rho$ corresponds to the atomic density of the material.

The reduced pair distribution function for sufficiently small $r$, i.e., $r \leq r_0$, is linear and proportional to the density, following:

\begin{equation}
\label{eq:sum_rule_R}
G(r \leq r_0) =  -4\pi\rho r
\end{equation}

The $Q$-dependent and dispersion-corrected form factor is represented by \cite{Wilson1993, Henke1993}:
\begin{equation}
\label{eq:formFactor}
f(Q) = f_\text{0}(Q) + \Delta f' + i\Delta f'',
\end{equation}

\noindent
where $\Delta f'$ and $\Delta f''$ correspond respectively to the anomalous and absorption corrections of Brennan and Cowan~\cite{Brennan1992}, while $f_0(Q)$ is the parameterized atomic form factor of Waasmaier and Kirfel \cite{Waasmaier1995}:
\begin{equation}
\label{eq:nthformFactor}
f_\text{0}(Q) = \sum_{j=1}^{5} a_\text{j}e^{-b_\text{j}\left(\frac{Q}{4\pi}\right)^2} + c,
\end{equation}
\noindent
in which $a$, $b$, $c$, are the parametrization coefficients of the form factor of a particular atom.

For a homogeneous alloy, one can approximate $ \langle f^2 \rangle$ and $\langle f \rangle^2$ as a mole fraction weighted sum of its constituent species, following:
\begin{equation}
\label{eq:formFactor_av_sq}
\langle f^2 \rangle = \sum\limits_{i=1}^n c_i f_{i}^2,
\end{equation}
and:
\begin{equation}
\label{eq:formFactor_sq_av}
\langle f \rangle^2 = \left( \sum\limits_{i=1}^n c_i f_{i} \right)^2,
\end{equation}
\noindent
whereby $n$ is the number of elements in the alloy and  $c_i$ is the mole fraction of element $i$.

The Compton intensities presented here were based on the tables of Cromer and Mann, \cite{10.1063/1.1712213, 10.1063/1.1670980}, which by Balyuzi \cite{Balyuzi:a11768} were parameterized as:
\begin{equation}
\label{eq:Compton}
(I_{\text{com}})_i  = Z_i^2 - \left( \sum_{j=1}^{5} a_\text{j,i}e^{-b_\text{j,i}\left(\frac{Q}{4\pi}\right)^2} + c_\text{i} \right),
\end{equation}
\noindent
with $Z_i$ being the atomic number of element $i$. The Compton intensity of the layers was approximated as a mole fraction weighted sum of $(I_{\text{com}})_i $, together with the characteristic wavelength shift of the Compton spectrum, according to:
\begin{equation}
\label{eq:Compton_weight}
I_{\text{com}} (Q) = \left( \frac{\lambda}{\lambda'}\right)^2 \sum\limits_{i=1}^n c_i(I_{\text{com}})_i,
\end{equation}
\noindent
where the characteristic wavelength shift, adopted from ~\cite{Thijsse}, can be computed as:
\begin{equation}
\label{eq:Compton_shift}
\lambda' = \lambda + \frac{2h}{m_\text{e}c}\sin^2(\theta),
\end{equation}
\noindent
in which $h$ is Planck's constant, $m_\text{e}$ is the electron mass and $c$ is the speed of light.

The linear attenuation coefficient can be constructed as:
\begin{equation}
\label{eq:mu}
\mu = 2r_e\rho \lambda \sum\limits_{i=1}^n c_i \Delta f''_i,
\end{equation}
where $r_e$ is the electron radius.


The polarization correction is:
\begin{equation}
P(\omega, \beta) = \begin{cases}
\frac{1 + \cos^2(2\alpha)\cos^2(\omega + \beta) }{1 + \cos^2(2\alpha)}, $ Source side [ref.~\cite{Hajdu}]$, \\ \\
\frac{1 + \cos^2(2\alpha)\cos^2(\omega + \beta) }{2}, $ Detector side [ref.~\cite{Thijsse}]$,
\end{cases}
   \label{eq:Polarization}
\end{equation}
\noindent
where $\alpha$ is the angle of the monochromator and with $\cos^2(2\alpha) = 1$ if no monochromator is being used. 

If one takes overillumination into account, the geometrical aberration, discussed by Rowles and coworkers~\cite{Rowles} and denoted $C$ here, can be summarized as:
\begin{equation}
C = \frac{d}{J} \frac{1}{\sin(\beta)},
   \label{eq:Geo}
\end{equation}
\noindent
where $d$ is the detector slit width and  $J$ is the illuminated footprint encompassed within the largest projected detector opening $d_{\text{max}}$. If the detector slit, or the pixels in the detector image as presented in this work, can be constructed to follow:
\begin{equation}
d(\beta) = d_{\text{max}}\sin(\beta),
   \label{eq:Geo}
\end{equation}
\noindent
then $C=1$, and no correction is needed.

Adopted from ~\cite{Thijsse}, the Lanczos filter applied to the structure factor takes the form:
\begin{equation}
\Tilde{F}(Q) = F(Q)\frac{ \sin( \pi Q/Q_{\text{max}}) }{ \pi Q/Q_{\text{max}} },
 \label{eq:Landczos}
\end{equation}
%


\section{Bulk SiO$_2$}\label{Appendix:SiO2}

\begin{figure}
	\centering
	\includegraphics[width= 0.8\columnwidth]{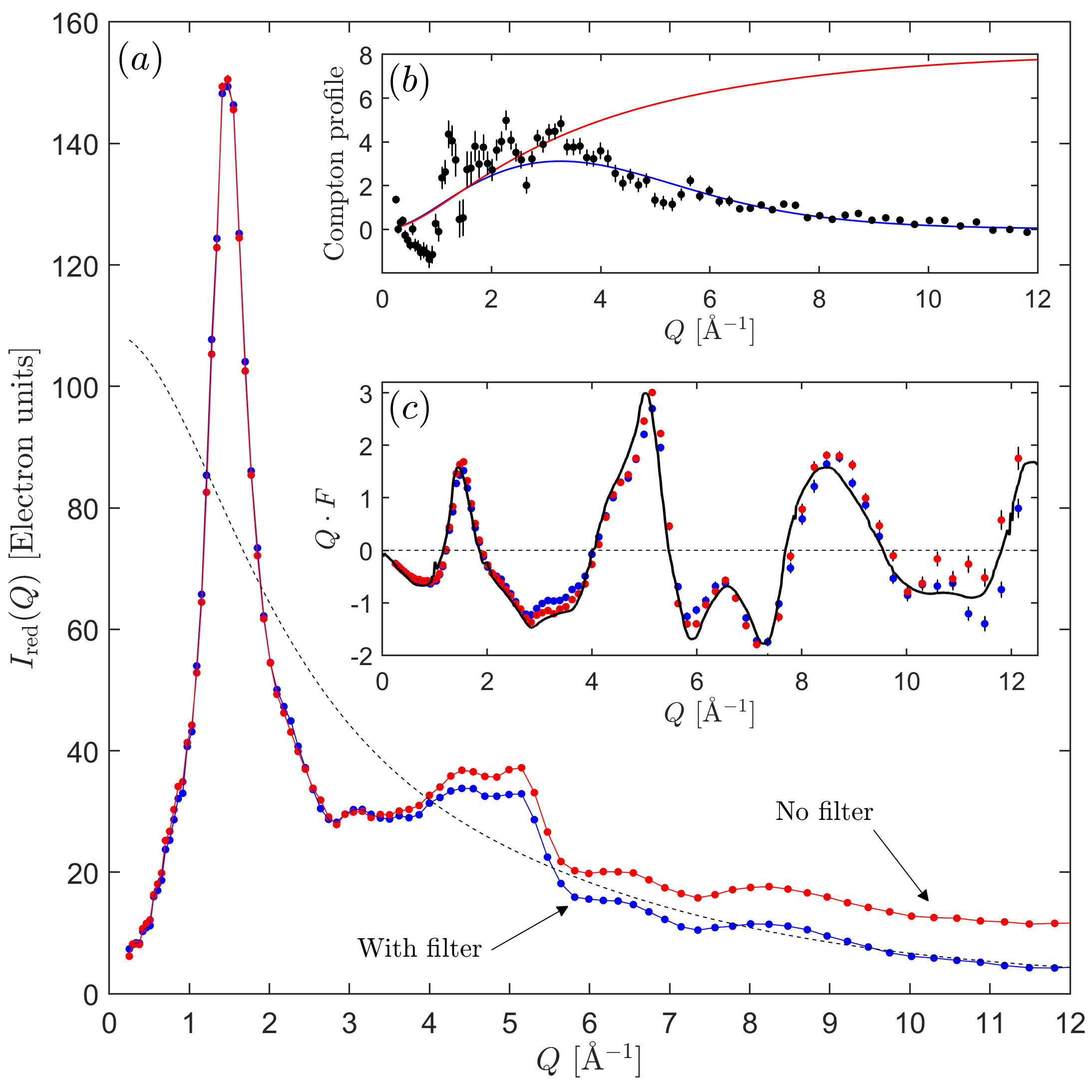}
	\caption{(a) The normalized scattering intensity in electron units ($I_{\text{red}} = N_0I_{\text{meas}}/(PA_{\text{f}})$) of bulk SiO$_2$ with energy filter (blue), and without energy filter (red). \\ 
		(b) The respective Compton profiles of the two scans, with the resulting explicit filtered profile in black, determined via their difference $I_{\text{com, filter}} = I_{\text{red, filter}} - (I_{\text{red, No filter}} - I_{\text{com, No filter}})$ . \\ 
		(c) The structure factors of the respective scans, with the traced structure factor data from Warren (Mozzi \& Warren, 1969), represented by the solid black line.}
	\label{fig:siO2}
\end{figure}

Figure \ref{fig:siO2}(a) depicts two GI scans of bulk SiO$_2$ following the outlined procedure of the article. The red data set was measured without an energy filter, while the blue data was measured with an energy filter. The inset, Figure \ref{fig:siO2}(b), show their respective Compton profiles, along with the explicit and error function fitted filtered Compton intensity, determined by taking the difference of the two scans, i.e., $I_{\text{com, filter}} =  I_{\text{red, filter}} - (I_{\text{red, No filter}} - I_{\text{com, No filter}})$. In the second inset, Figure \ref{fig:siO2}(c), the structure factors of the respective scan, together with the structure factor adapted from Warren \cite{Mozzi1969TheSO} are displayed, highlighting that the GI implementation discussed in this article is also applicable to bulk samples.

\ack{The authors gratefully acknowledge a productive partnership with Malvern Panalytical. The authors further gratefully acknowledge the Swedish Research Council (VR -  Vetenskapsrådet) grant number 2018-05200 for funding the work and the Swedish Energy Agency grant 2020-005212. G.K.P. also acknowledges the Carl-Tryggers foundation grants CTS 17:350 and CTS 19:272 for financial support. The computations were enabled by resources provided by the Swedish National Infrastructure for Computing (SNIC), partially funded by the Swedish Research Council through grant agreement no.\ 2018-05973.}


\bibliographystyle{iucr}
\bibliography{iucr}

\end{document}